\documentclass{moriond}
\usepackage{wrapfig}

\def\be{\begin{equation}}
\def\ee{\end{equation}}
\def\bea{\begin{eqnarray}}
\def\eea{\end{eqnarray}}

\newcommand{\Photo}{\includegraphics[height=35mm]{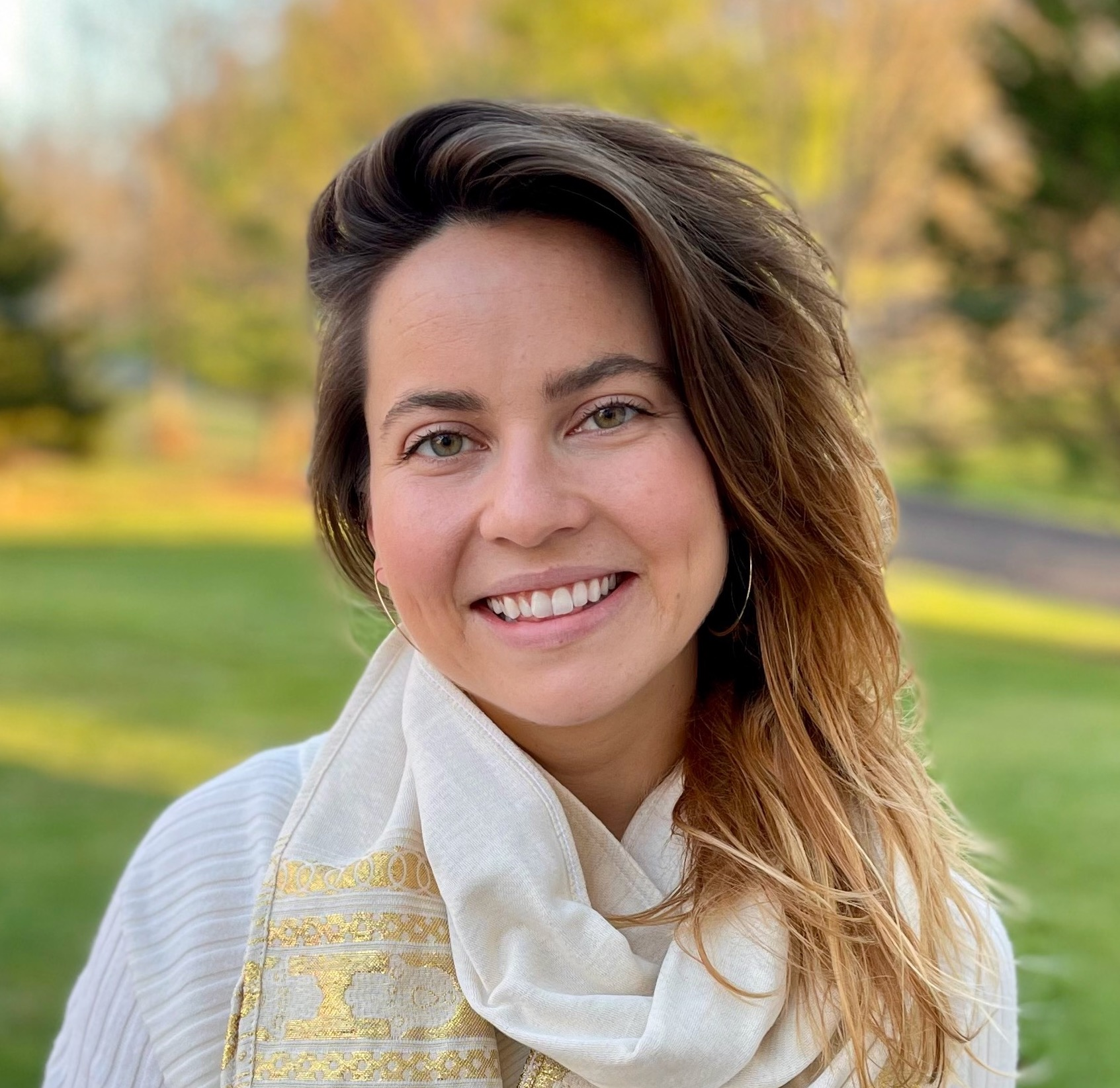}}

\begin{document}
\vspace*{4cm}
\title{Highlights from Long-Lived Particle Searches at ATLAS}

\author{Julia L. Gonski, on behalf of the ATLAS Collaboration}

\address{Nevis Laboratories, Columbia University\\
136 S Broadway, Irvington, NY 10533}

\maketitle\abstracts{
The latest results of long-lived particle (LLP) searches from the ATLAS Experiment at the Large Hadron Collider are presented.
Analyses are presented with a focus on detector subsystem needed to discern the LLP signature from Standard Model background, and the custom reconstruction requirements for sensitivity.
Results are contextualized with updated summary plots for key new physics candidates, along with notes for future LLP searches. 
}

\section{Introduction}
\label{sec:intro}

\footnotetext{Copyright 2022 CERN for the benefit of the ATLAS Collaboration. CC-BY-4.0 license.}

The search program for beyond the Standard Model physics with the Large Hadron Collider (LHC) is designed to cover new particle hypotheses that are theoretically well-motivated, discernible with collider detector technologies, and in phase space that is not well-covered by other experiments.
Long-lived particles (LLPs) satisfy all of the above criteria, where the lifetime can naturally emerge from reduced phase space for a decay due to nearly degenerate mass spectra, or small couplings between parent and daughter particles.
Existing mass limits for new physics are in many cases invalid if the candidate has a lifetime that is non-prompt on the scale of LHC detectors, making these searches crucial for thorough coverage of BSM possibilities. 
However, the uniqueness of LLP signatures 
requires the novel and highly customized use of ATLAS~\cite{atlas} detector signals for a sensitive analysis. 
The location of an LLP decay in the detector depends on its lifetime, which can theoretically span many orders of magnitude, thus ideally the decay is reconstructible at any radial value within the detector.
Recent results from ATLAS LLP searches are presented here in radial order of the subsystem that is used to provide its signal.

\section{New Search Results}
\label{sec:searches}

\subsection{Displaced Heavy Neutral Leptons}
\label{subsec:dHNL}

A search is presented for long-lived and thus displaced heavy neutral leptons (dHNLs)~\cite{dhnl}.
This signature emerges from a massive right-handed sterile neutrino that can potentially explain the neutrino mass hierarchy and baryon asymmetry. 
The dHNL $\mathcal{N}$ is produced in the weak interaction $W \rightarrow \mathcal{N} \ell$, and the $\mathcal{N}$ then decays with some lifetime into two oppositely-charged leptons and a neutrino. 
The final state of this process is one prompt lepton, used to trigger, and an opposite sign dilepton displaced vertex.
The $\mathcal{N}$ mass is reconstructed using the four vectors of the daughter leptons and neutrino.
Displaced lepton efficiency is aided by the use of \textit{large-radius tracking} (LRT), an additional algorithm that runs over inner detector hits not included in standard tracks.
This allows for the recognition of hit patterns that do not necessarily coincide with the beamline, enhancing efficiency for displaced tracks and thus displaced leptons.
\begin{wrapfigure}[25]{r}{0.38\textwidth}
    \includegraphics[width=0.38\textwidth]{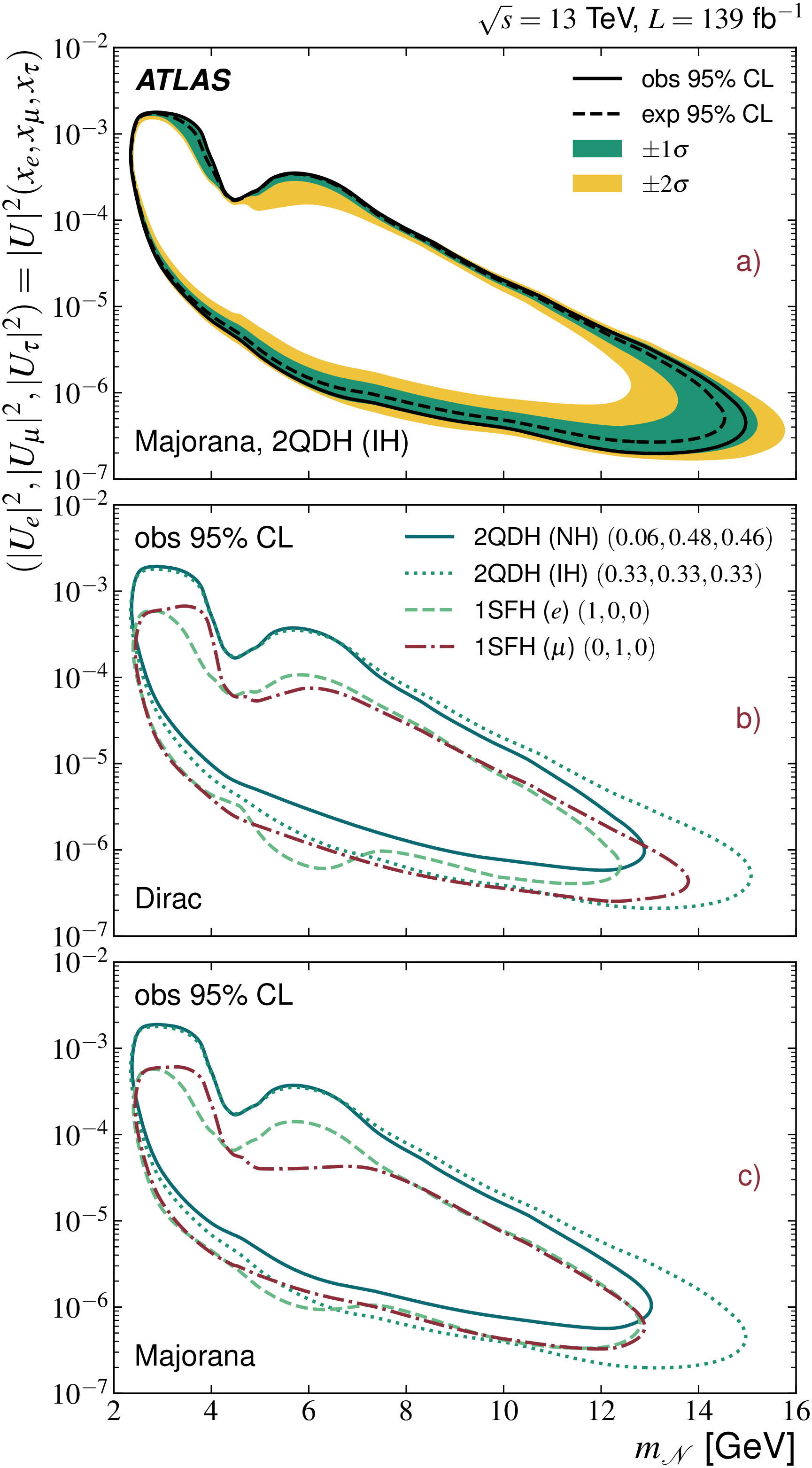}
    \caption{\label{fig:dHNL_fig_02} 95\% confidence limits on $|U|^2$ vs. $m_\mathcal{N}$ in the Majorana case.}
\end{wrapfigure}
The background to this process is dominated by random track crossings, 
and is estimated in the signal region (SR) ($m_{\mathcal{N}} < $ 20 GeV) with a fully data-driven process of generating toy m$_{\mathcal{N}}$ values from the shuffling of leptons from control region (CR) events ($m_{\mathcal{N}} >$  20 GeV).

No signal is observed in the unblinded SR data, and the results are interpreted in the weak-like dimensionless mixing angles $|U_{\alpha}|^2$, where $\alpha$ refers to the lepton flavor, for both Majorana and Dirac mass scenarios.
Figure~\ref{fig:dHNL_fig_02} shows the resulting 95\% CL exclusion contours of $|U|^2$ as a function of $m_\mathcal{N}$.
This result provides an improvement of the existing limit on $|U_\mu|^2$, the first limits from ATLAS on $|U_e|^2$, and the first limits on multi-flavor mixing scenarios motivated by neutrino flavor oscillations.

\subsection{Pixel $dE/dx$}
\label{subsec:pixeldEdx}

A search is presented for events with tracks that have anomalously high energy deposition ($dE/dx$) in the inner detector, which come from a new heavy charged particle with a relativistic $\beta$ that is measurably less than 1~\cite{pixel}.
Innovative use of $dE/dx$ measurements in the pixel detector allow for reconstruction of the track mass through momentum measurement and parameterization of the Bethe-Bloch equation.
A track in the pixel detector is built from clusters of hits, and its $dE/dx$ value is the average of the ionization measurement across all clusters in the track.
The value of the crucial analysis variable $dE/dx$$_\mathrm{correct}$ is obtained from $dE/dx$ through a two step procedure. 
First, the highest energy clusters are removed from the track, to mitigate the effect of the Landau tail.
Then the relationship with respect to particle $\beta\gamma$ is calibrated to known mass examples from the SM using data from low pileup runs.
\begin{wrapfigure}[13]{l}{0.35\textwidth}
    \centering
    \includegraphics[width=0.35\textwidth]{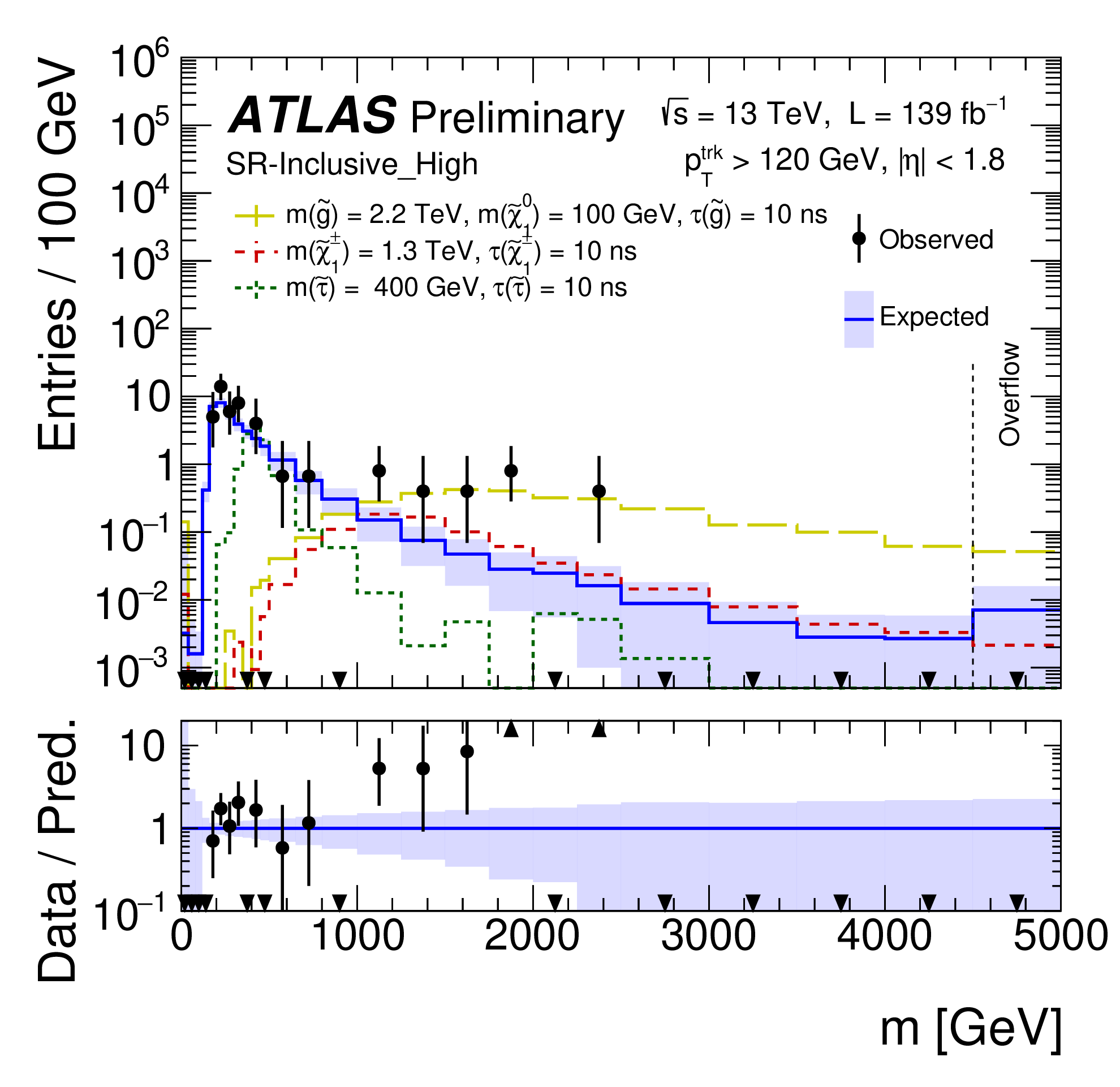}
    \caption{\label{fig:pixel_fig_15} The observed mass distribution in an analysis SR, along with expected background and signals.}
\end{wrapfigure}

SRs for the search are built by selecting for central ($|\eta| <$ 1.8) isolated tracks with high $p_\mathrm{T}$ ($>$ 120 GeV) and large $dE/dx$$_\mathrm{correct}$ ($>$ 1.8 MeV g$^{-1}$ cm$^2$), along with high $E_\mathrm{T}^\mathrm{miss}$ ($>$ 170 GeV) from the lightest supersymmetric particle (LSP) in the final state.
The background estimation is toy-based, where the track mass spectrum in the SR is obtained by drawing a value of (1/$p_\mathrm{T}$, $\eta$) from the kinematic CR, selecting its $dE/dx$ value according to its $\eta$ from the $dE/dx$ CR, and finally calculating track mass using momentum and $\beta\gamma$.
The resulting toy mass distribution is validated to data in orthogonal selections with low track $p_\mathrm{T}$ and high track $\eta$.

Unblinded distributions of data and expected background can be seen in Figure~\ref{fig:pixel_fig_15}.
The largest excess is observed in $dE/dx$ $>$ 2.4 MeV g$^{-1}$ cm$^2$ corresponding to 3.6 (3.3) $\sigma$ local (global) significance.
A cross check of candidate tracks with time-of-flight variables from the tile calorimeter and muon spectrometer was consistent with $\beta$ = 1, therefore not consistent with the LLP hypothesis.
Results are interpreted spanning a variety of masses and lifetimes for new SUSY particles, as shown on the summary plot in Figure~\ref{fig:summary}.

\subsection{Non-Pointing Photons}
\label{subsec:npp}

\begin{wrapfigure}[14]{r}{0.42\textwidth}
    \centering
    \includegraphics[width=0.42\textwidth]{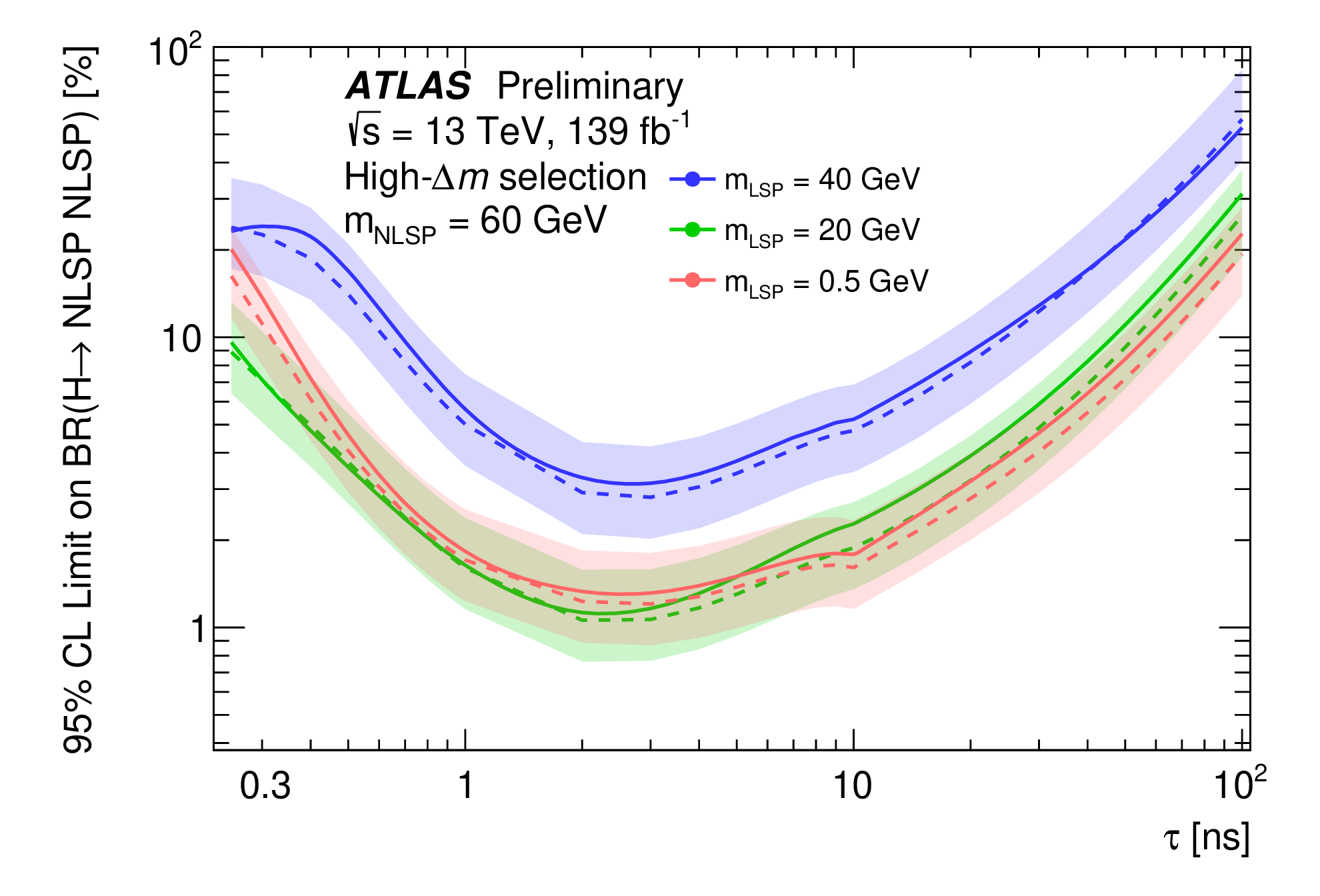}
    \caption{\label{fig:npp_fig_09a} 95\% confidence level limits on the Higgs to two NLSP BR as a function of NLSP lifetime.}
\end{wrapfigure}
A search is presented for photons which come from the decay of an LLP and thus are late with respect to the bunch crossing (delayed) and do not point back to the primary interaction point on the beamline (\textit{non-pointing})~\cite{npp}.
The process studied here is the decay of a Higgs boson to two next-to-lightest supersymmetric particles (NLSP), followed by the NLSP decay to an LSP and a photon.
The final state photons can be separated from background using precise information from the liquid argon (LAr) electromagnetic calorimeter, which can provide a timing measurement with $\sim$200 ps of resolution for energy deposits above $\sim$10 GeV.
Its segmentation also allows for the computation of \textit{pointing}, which is defined as the distance in $z$ between the primary vertex and the extrapolation of the photon trajectory back to the beamline.

The SR is constructed to select events with photons that have high timing, high pointing, and $E_\mathrm{T}^\mathrm{miss}$ from the final state LSPs.
Further selections are made that separate the SR into high and low bins of the NLSP to LSP mass splitting, which dictates the final state photon kinematics.
The background estimation is fully data-driven due to the non-Gaussian tails in these sensitive analysis variables.
The timing shape in the SR is predicted from an admixture of two templates taken from CRs, one enriched in real photons and one enriched in fake photons.

No excess of data over the estimated background is found in the SR.
As a result, limits are set on the branching ratio (BR) of the Higgs boson to two NLSP process.
An example is shown in Figure~\ref{fig:npp_fig_09a}, as a function of the LSP lifetime.
This is the most sensitive analysis region, with high mass splitting, allowing for a BR exclusion down to 1\%.

\subsection{Displaced \texttt{CalRatio} Jets}
\label{subsec:calratio}

A search is presented for displaced jets that come from $s$-channel production of an exotic hidden sector scalar mediator, which decays into two LLPs that decay to fermions~\cite{calratio}.
The signature in the detector is that of jets that are narrow, trackless, and have an anomalously low ratio of electromagnetic to hadronic calorimeter energy deposits (so-called \texttt{CalRatio}) jets.
Signal jets are selected based on this energy ratio, along with the outputs from a jet-level neural net (NN) and event-level boosted decision tree to reject cosmic rays, beam-induced background, and SM multijet processes.
A novel adversary is added to the neural net training to remove the effect of mismodeling in QCD simulation.
The observed data in the SR is found to be compatible with the estimated background.
The theoretical interpretation covers mediator masses between 60 GeV and 1 TeV,  including the case in which the mediator is the 125 GeV Higgs boson, and LLP masses between 5 and 475 GeV.
Excluded phase space from this search is shown in the context of other recent hidden sector analyses in Figure~\ref{fig:summary}.

\subsection{Dark Photon Lepton-Jets}
\label{subsec:darkphotons}

A search is presented for highly collimated SM fermions known as a \textit{lepton-jet} that come from the decay of a light neutral LLP that is a dark photon candidate~\cite{darkphotons}.
Events are selected through custom reconstruction algorithms which distinguish between muonic and calorimeter dark photon jets (DPJs).
Muonic DPJs are reconstructed by clustering of MS-only tracks, with a deep NN trained over low-level detector info to distinguish them from cosmic ray backgrounds.
Calorimeter DPJs select jets with a low ECAL energy fraction and employ a convolutional NN trained with $\eta$/$\phi$ maps for background rejection. 
The background estimation is fully data-driven using the ABCD method over variables unique to the two different flavor channels.
No excess in the data is found, 
and the first ATLAS exclusions are set for electron flavor dark photon decays, as well as for dark photon masses below 0.1 GeV.

\section{Conclusions}
\label{sec:conclusions}

\begin{figure}[h!]
    \centering
    \includegraphics[width=0.48\textwidth]{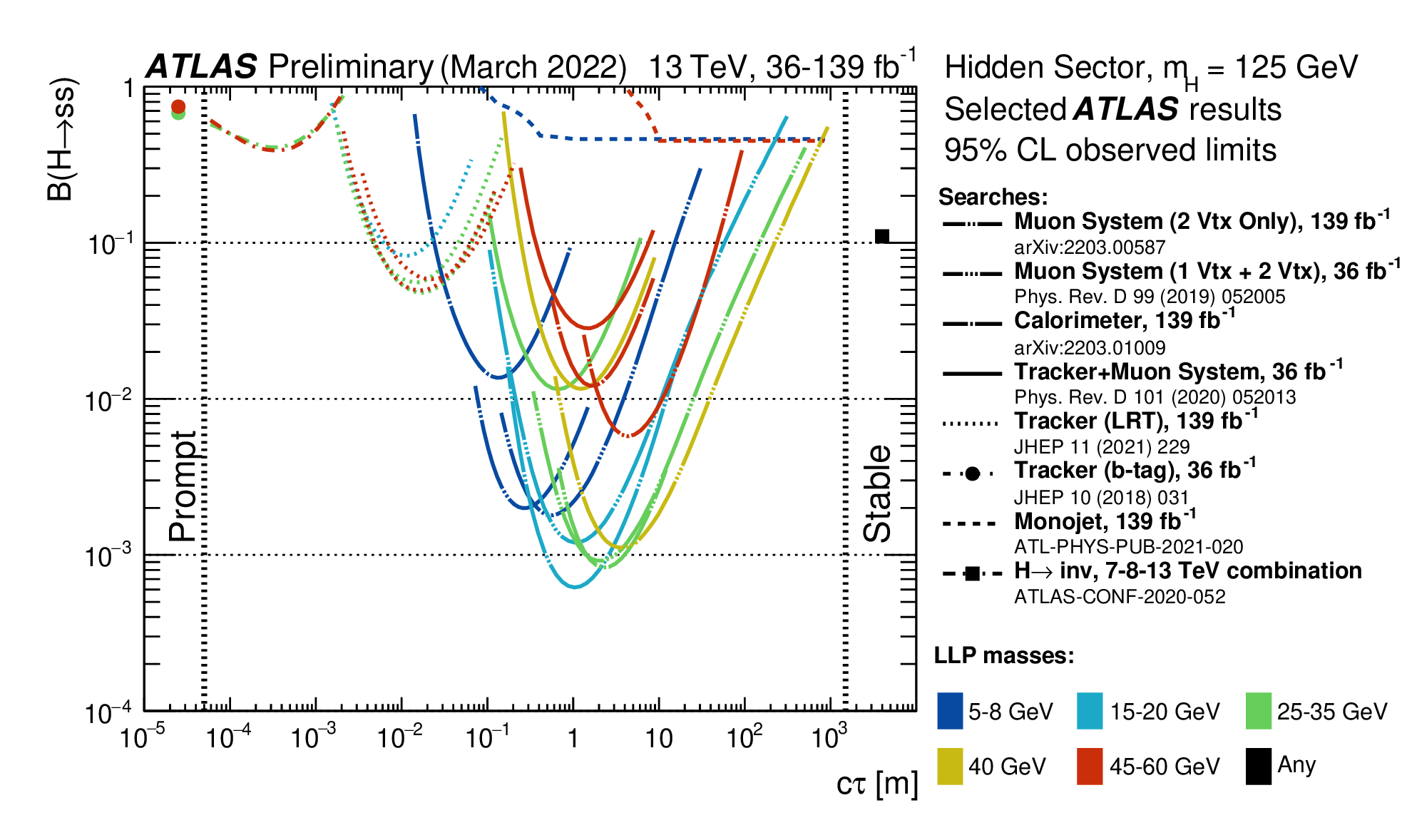}
    \includegraphics[width=0.3\textwidth]{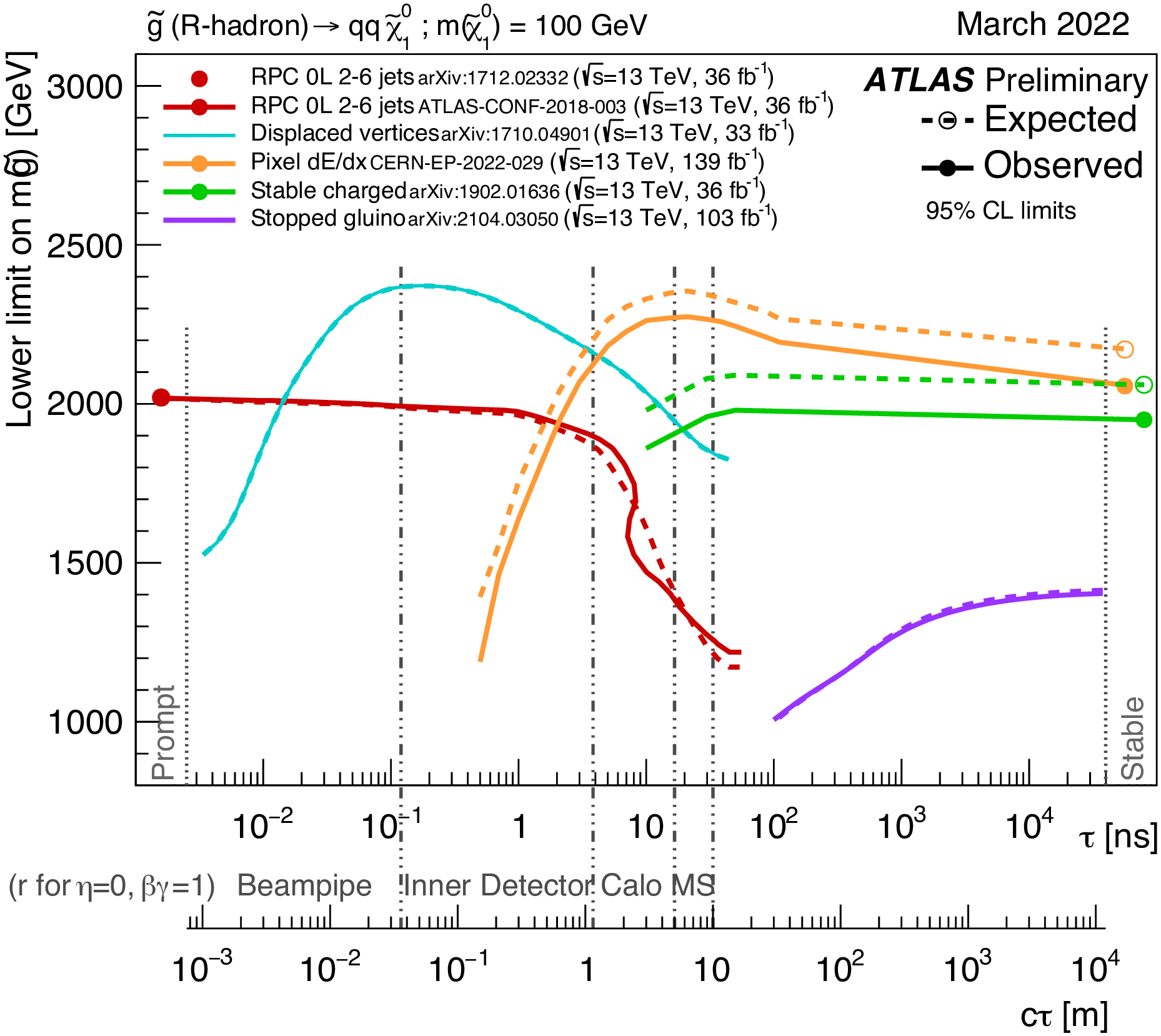}
    \caption{\label{fig:summary} 95\% confidence limits on the Higgs BR as a function of $s$ $c\tau$ for a $H\rightarrow ss$ Hidden Sector process (left), and constraints on the gluino mass-vs-lifetime plane, where contours from the pixel $dE/dx$ and stable charged particle search are extrapolated to the stable regime with a straight line (right).}
\end{figure}

Several new results from LLP searches at ATLAS have been presented.
They represent both the improvement of existing limits on key new physics candidates, as well as several first results in previously unexplored phase space.
Summary plots of hidden sector (SUSY) searches have been updated to include the displaced \texttt{CalRatio} jet (pixel $dE/dx$) result, shown in Figure~\ref{fig:summary}. 


The landscape for ATLAS LLP searches will expand starting with LHC Run 3 data-taking in 2022.
The large-radius tracking algorithm introduced in Section~\ref{subsec:dHNL} will be used by default in all future data reconstruction.
Additionally, it will be integrated into the ATLAS high-level trigger (HLT) chain, allowing for the construction of new displaced lepton triggers that run over the LRT track collection and can provide significant efficiency gains, particularly for lower $p_\mathrm{T}$ objects.
These advances, coupled with the continuing exploration of theoretically interesting phase space, will continue to drive the importance and efficacy of the ATLAS LLP search program.



\section*{Acknowledgments}

The author is supported by the National Science Foundation under Grant No. PHY-2013070.


\end{document}